\documentclass[npg]{copernicus}
\newcommand{\D}{\mathrm d}
\newcommand{\E}{\mathrm e}
\newcommand{\I}{\mathrm i}

\begin{document}

%\linenumbers

\title{On the applicability of the Hasselmann kinetic equation to the Phillips spectrum.}

\author[1,2]{Alexander\,O.\,Korotkevich}
\author[3,4,5,2]{Vladimir\,E.\,Zakharov}

\affil[1]{Department of Mathematics \& Statistics, The University of New Mexico, MSC01 1115, 1 University of New Mexico, Albuquerque, New Mexico, 87131-0001, USA}
\affil[2]{L.\,D.~Landau Institute for Theoretical Physics, 2 Kosygin Str., Moscow, 119334, Russian Federation}
\affil[3]{University of Arizona, Department of Mathematics, Tucson, AZ 85721, USA}
\affil[4]{P.\,N.~Lebedev Physical Institute, 53 Leninskiy prospekt, Moscow, 119334, Russian Federation}
\affil[5]{Laboratory of Nonlinear Wave Processes, Novosibirsk State University, Novosibirsk, Russian Federation}

%% The [] brackets identify the author to the corresponding affiliation, 1, 2, 3, etc. should be inserted.

\runningtitle{On the applicability of the kinetic equation...}

\runningauthor{Korotkevich, A.\,O., Zakharov, V.\,E.}

\correspondence{A.\,O.\,Korotkevich\\ (alexkor@math.unm.edu)}

\received{}
\pubdiscuss{} %% only important for two-stage journals
\revised{}
\accepted{}
\published{}

%% These dates will be inserted by the Publication Production Office during the typesetting process.

\firstpage{1}

\maketitle  %% Please note that for the copernicus2.cls this command needs to be inserted after \abstract{TEXT}

\begin{abstract}
We investigate applicability of the Hasselmann kinetic equation to the spectrum of surface gravity waves at different levels
of nonlinearity in the system, which is measured as average steepness. It is shown that even in the case of relatively
high average steepness, when Phillips spectrum is present in the system, the spectral lines are still very narrow,
at least in the region of direct cascade spectrum. It allows us to state that even in the case of Phillips spectrum
the kinetic equation can be applied to the description of the ensembles of ocean waves.
\end{abstract}

\introduction  %% \introduction[modified heading if necessary]
Two really seminal papers were published in the area of Physical Oceanography about half a century ago. We mean the article
by~\citet{Phillips1958} and the work of~\citet{Hasselmann1962}. Both of them were concentrated on the same problem: what is
going on with ocean waves growing under influence of wind?

O.\,Phillips suggested that this growth is arrested by wave-breaking,
in other words by formation of ``white caps'' (or ``white horses''). Wave breaking is the main mechanism of energy dissipation.
In ``white caps'' mechanical energy of waves transforms to turbulence on small scales, then to heat. This is a strongly
nonlinear phenomenon which cannot be studied by perturbative methods. An analytical theory of wave breaking is not developed
yet. The Phillips' assumption on predomination of wave-breaking effects made him possible to offer, on the base of
dimensional considerations, the universal energy spectrum of wind driven waves, so called Phillips' spectrum.

K.\,Hasselmann developed a completely different theory. He noticed that a typical ensemble of ocean waves hoards a small
parameter -- average steepness (slope) of the surface $\mu$. One can define it, for instance, as
$\mu=\sqrt{\langle|\nabla\eta|^2\rangle}$, where $\eta$ is the shape of the surface. This definition has the most clear
geometrical meaning. Another definition estimates it as average amplitude $A$ of the waves multiplied by wave-vector of the
 spectral peak $k_p$. This definition is easy to use for experimental observations. Later we shall consider another
definition, which involves spectral distribution function. All these definitions give very close values of the average
steepness. Typically $\mu\simeq 0.1$. This
fact allows to assume that the wind-driven sea is a community of weakly interacting waves, which can be described statistically
by the use of expansion in powers of average steepness $\mu$. This is a tedious procedure because one have to expand up to order
$\mu^4$. However K.\,Hasselmann coped with this hard work and derived his famous Hasselmann kinetic equation for squared
wave amplitudes.

Then it was found  that Hasselmann equation has exact stationary solutions (Kolmogorov-Zakharov or KZ
spectra)~\citep{ZF1967,ZF1967JAMTP,ZZ1982,ZLF1992} decaying at high frequency region slower than Phillips spectrum.
A question arises which theory is more correct? The answer is given by experiment. Numerous measurements made
in lakes and ocean showed that the real spectrum is a combination of weak-turbulent KZ spectra and the Phillips spectrum.
If $\omega_p$ is spectrum peak frequency, in the energy containing range $\omega_p < \omega < 3\omega_p$ KZ spectra
are realized, while the Phillips spectrum predominates in the high-frequency range. This fact has the following explanation.
Wave turbulence in the wind driven sea is a mixture of weak and strong turbulence. This is a question of phase correlation.
Weakly nonlinear interaction provides correlation of waves phases on large scales -- hundreds of characteristic
wavelengths. At the same time the wave breaking and white capping is the localized phenomenon. As a result looking
at the wind driven sea one observes formation of short living localized wave breaking events embedded into homogeneous
weakly nonlinear background.

Similar situation, coexistence of weak turbulence and localized coherent structures, is typical for wave turbulence.
It takes place, for instance, in nonlinear optics~\citep{DNPZ1992} where optical turbulence coexists with self-focusing
or in isotropic plasma where weak Langmuir turbulence coexists with Langmuir collapses~\citep{ZLF1992}. Similar phenomena
were observed recently in numerical simulations for gravity waves~\citep{ZKPR2007}. In all these situations the usual course
of action is to augment corresponding kinetic equations of weak turbulence by introduction of an additional
term, describing dissipation of energy in the coherent structures (corresponding to white capping onsets)~\citep{ZKP2009}.
The question is it still reasonable to
use kinetic equation as an adequate model of waves interaction in the presence of coherent structures?

In this paper we give positive answer to this question. To do this we perform massive numerical simulation of
primordial dynamical equations describing potential flow of Euler equations of ideal fluid with a free surface.
In our numerical experiments we studied Fourier spectra of space-time correlation function of normal canonical
variables describing the surface dynamics. Our goal was to determine the ``shape of the line'', i.e. frequency
spectra of certain spatial Fourier harmonics. According to the theory of weak turbulence these spectra must be narrow
and concentrated near the frequency given by the dispersion relation for waves of small amplitude. We obtained the following
important result -- in a broad range of wave numbers the spectra remain narrow even in the presence of wave-breaking
events, when the spatial spectrum obeys the Phillips law. This result opens a door to construction of a properly justified
dissipation term which can be used thereafter in wave prediction operational models.

\section{Basic model}
We solve numerically weakly nonlinear Euler equations for dynamics of incompressible deep fluid with free surface
in the presence of gravity by the pseudo-spectral code described in~\citet{KDZ2012}. The code was verified in our previous
papers~\citep{DKZ2003cap,DKZ2003grav,DKZ2004,ZKPD2005,ZKPR2007,KPRZ2008,Korotkevich2008PRL,Korotkevich2012MCS}. The equations
are written for surface elevation $\eta(x,y,t)$ and hydrodynamic velocity potential on the surface $\psi(x,y,t)$. Equations
are result of weakly nonlinear expansion of the Hamiltonian up to the fourth order terms in steepness~\citep{ZLF1992}
\begin{equation}
\label{eta_psi_system}
\begin{array}{rl}
\displaystyle
\dot \eta = &\hat k \psi - (\nabla (\eta \nabla \psi)) - \hat k [\eta
\hat k \psi] + \hat k (\eta \hat k [\eta \hat k \psi])\\
\displaystyle
 &+
\frac{1}{2} \Delta [\eta^2 \hat k \psi] + \frac{1}{2} \hat k [\eta^2
\Delta\psi] - \widehat F^{-1} [\gamma_k \eta_k],\\
\displaystyle
\dot \psi
= &- g\eta - \frac{1}{2}\left[ (\nabla \psi)^2 - (\hat k \psi)^2
\right] - [\hat k \psi] \hat k [\eta \hat k\psi]\\
\displaystyle
&- [\eta \hat k \psi]\Delta\psi - \widehat F^{-1} [\gamma_k
\psi_k] + \widehat F^{-1} [P_{\vec k}].
\end{array}
\end{equation}
Here dot means time-derivative, $\Delta$ -- Laplace operator in $\vec r = (x,y)^{T}$-plane, $\hat k = \sqrt{-\Delta}$,
$\widehat F^{-1}$ is an inverse Fourier
transform, $\gamma_k$ is a dissipation rate (according to recent
work by~\citet{DDZ2008} it has to be included in both equations), which
corresponds to viscosity on small scales and, if needed, ``artificial''
damping on large scales.  $P_{\vec k}$ is the driving term which
simulates pumping on relatively large scales (for example, due to wind). Here and further we use
symmetric Fourier transform
$$
\eta_{\vec k} = \frac{1}{2\pi} \int \eta_{\vec r} \E^{-\I {\vec k} {\vec r}} \D^2 r.
$$
Normal variables are introduced as follows
\begin{equation}
\eta_{\vec k} = \sqrt{\frac{k}{2\omega_k}}(a_{\vec k} + a_{-\vec k}^{*}),\;\;\;
\psi_{\vec k} = \I\sqrt{\frac{\omega_k}{2 k}}(a_{\vec k} - a_{-\vec k}^{*})
\end{equation}
where $\omega_k = \sqrt{gk}$, $g$ -- gravity acceleration which was taken $g=1$.
The equations were solved in a $2\pi\times 2\pi$ periodic
box. Number of modes was $1024\times 1024$. Inside the box spectra were practically isotropic, thus one can put
$$
|\eta_{\vec k}|^2\simeq \frac{k}{2\omega_k}(n_{\vec k} + n_{-\vec k}),\;\;\; n_{\vec k} = |a_{\vec k}|^2.
$$
We define the ``current steepness'' $\mu_k$ by relation
\begin{equation}
\mu_k = 2\pi\int\limits_{0}^{k} k^{7/2} n_k \D k.
\end{equation}
Here we use angle-averaged $n_k$ and summation is replaced by integration over $k=|\vec k|$. Average steepness
can be obtained as a limit:
$$
\mu = \lim\limits_{k\rightarrow +\infty} \mu_{k}.
$$
This definition gives a value of average steepness which differs from the geometrical definition
$\mu=\sqrt{\langle|\nabla\eta|^2\rangle}$ in our experiments at max by value close to $0.005$, which is just several
percent of the characteristic average steepness.

For the spatial harmonic with wave vector $\vec k$ we defined the time Fourier transform as follows
$$
a(\vec k,\omega) = \frac{1}{T}\int\limits_{0}^{T}a(\vec k, t)\E^{-\I\omega t}\D t.
$$
Here $T=2\pi/\omega_k$ -- period of the chosen wave. Thereafter $I(\vec k,\omega)=|a(\vec k,\omega)|^2$.
In equations~(\ref{eta_psi_system}) pumping had the form
\begin{equation}
P_{\vec k} = f_k \E^{\I R_{\vec k} (t)}, f_k = \cases{
4 F_0 \frac{(k-k_{p1})(k_{p2}-k)}{(k_{p2} - k_{p1})^2},\cr
0 - \mathrm{if}\; k < k_{p1}\;\mathrm{or}\; k > k_{p2};\cr}
\end{equation}
here $k_{p1}=28,\; k_{p2}=32$ and $F_0 = 1.5\times 10^{-5}$; $R_{\vec
k}(t)$ was a uniformly distributed random number in the interval
$(0,2\pi]$ for each $\vec k$ and $t$. The initial condition was low
amplitude noise in all harmonics. Time step was $\Delta t =
6.7\times 10^{-4}$. The dissipation function $\gamma_k = \gamma_k^{(1)}+\gamma_k^{(2)}$.

Artificial viscosity $\gamma_k^{(1)}$ was the same in all experiments
\begin{equation}
\gamma_{k}^{(1)} = \cases{
\gamma_0 (k - k_d)^2, \;k > k_d,\cr
0, \mathrm{if}\;k \le k_d;\cr}
\end{equation}
where $k_d = 256$ and $\gamma_0 = 0.97\times10^{2}$.
Due to presence of this dissipation the most part of our phase space, actually for $k>256$, was ``passive''.
But this was necessary in order to get rid of aliasing harmonics and to simulate dissipation of energy.

$\gamma_k^{(2)}$ was dissipation concentrated in small wave numbers. It was zero only in one of three experiments.

\section{Description of experiments}
We performed three series of experiments, choosing different functions $\gamma_k^{(2)}$, describing damping in the area
of small wave-numbers.
\subsection{Without condensate and inverse cascade}
\begin{figure}[tb]
%\vspace*{2mm}
\begin{center}
\includegraphics[width=8.3cm]{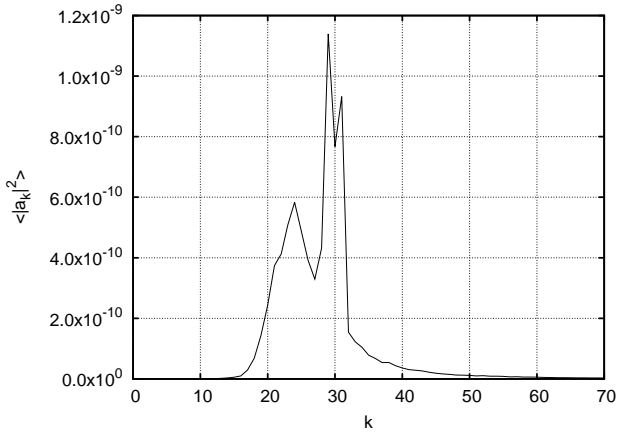}
\end{center}
\caption{\label{figure1}Spectrum with both condensate and inverse cascade suppressed. Linear scale.}
\end{figure}

\begin{figure}[tb]
%\vspace*{2mm}
\begin{center}
\includegraphics[width=8.3cm]{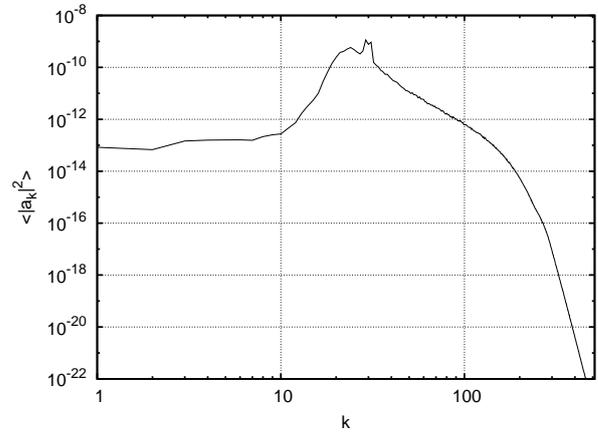}
\end{center}
\caption{\label{figure2}Spectrum with both condensate and inverse cascade suppressed. Logarithmic scale.}
\end{figure}

\begin{figure}[tb]
%\vspace*{2mm}
\begin{center}
\includegraphics[width=8.3cm]{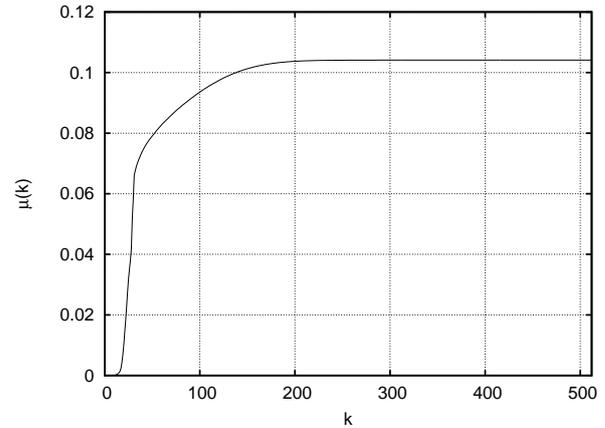}
\end{center}
\caption{\label{figure3}Steepness as a function of absolute vale of wave-number. Both condensate and inverse cascade were suppressed.}
\end{figure}

\begin{figure}[tb]
%\vspace*{2mm}
\begin{center}
\includegraphics[width=8.3cm]{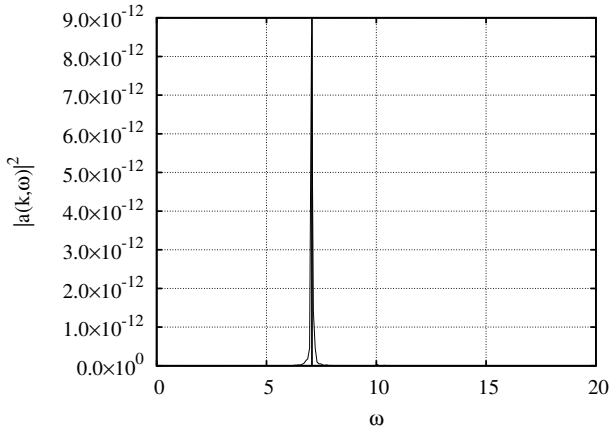}
\end{center}
\caption{\label{figure4}Spectral line of the harmonic $\vec k = (0, 50)^{T}$. Both condensate and inverse cascade were suppressed.}
\end{figure}

\begin{figure}[tb]
%\vspace*{2mm}
\begin{center}
\includegraphics[width=8.3cm]{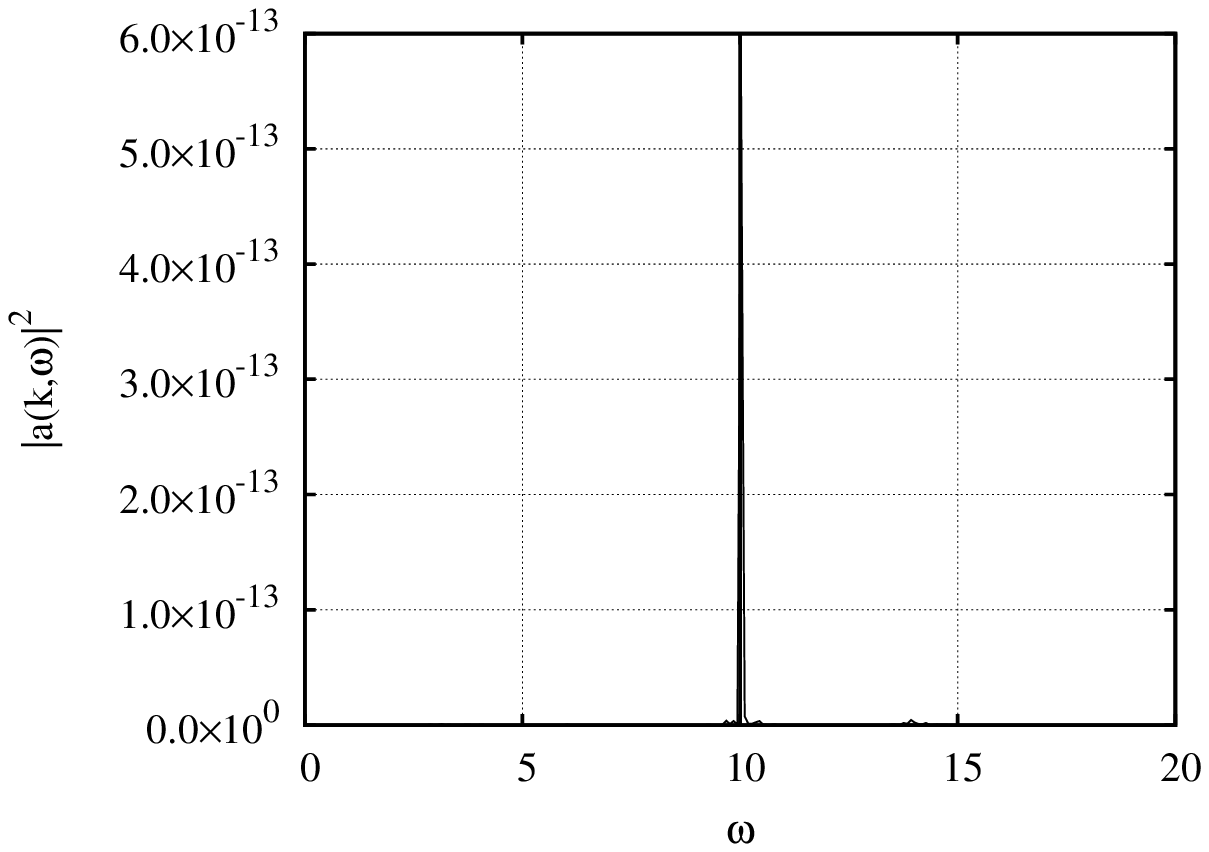}
\end{center}
\caption{\label{figure5}Spectral line of the harmonic $\vec k = (0, 100)^{T}$. Both condensate and inverse cascade were suppressed.}
\end{figure}

\begin{figure}[tb]
%\vspace*{2mm}
\begin{center}
\includegraphics[width=8.3cm]{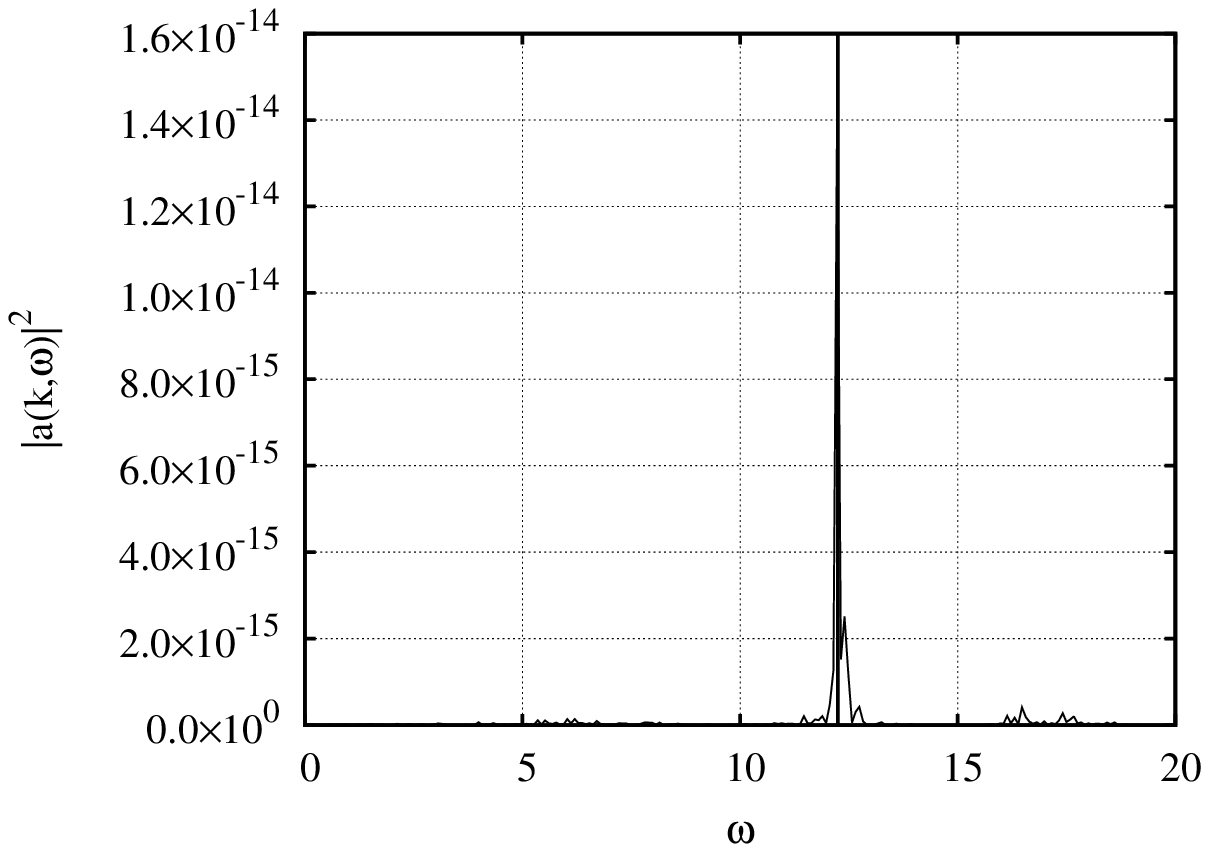}
\end{center}
\caption{\label{figure6}Spectral line of the harmonic $\vec k = (0, 150)^{T}$. Both condensate and inverse cascade were suppressed.}
\end{figure}

\begin{figure}[tb]
%\vspace*{2mm}
\begin{center}
\includegraphics[width=8.3cm]{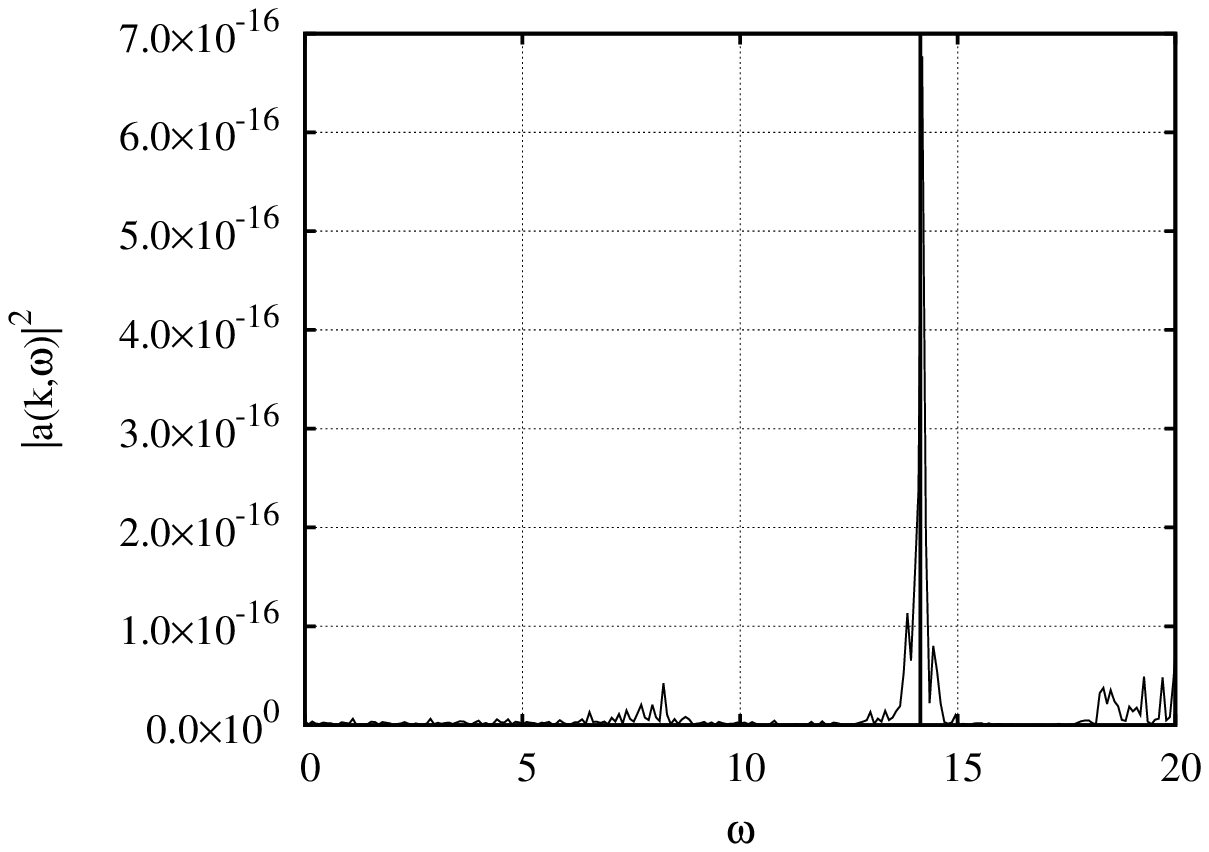}
\end{center}
\caption{\label{figure7}Spectral line of the harmonic $\vec k = (0, 200)^{T}$. Both condensate and inverse cascade were suppressed.}
\end{figure}

\begin{figure}[tb]
%\vspace*{2mm}
\begin{center}
\includegraphics[width=8.3cm]{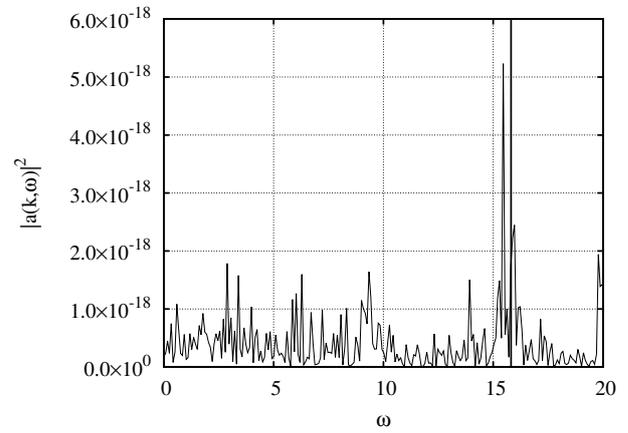}
\end{center}
\caption{\label{figure8}Spectral line of the harmonic $\vec k = (0, 250)^{T}$. Both condensate and inverse cascade were suppressed.}
\end{figure}

In the first series of experiments we assumed that it was linearly growing in small wave-numbers
$k<28$ and equal to
$$
\gamma_{k}^{(2)} = \cases{
0.2 |k - 28|, \;k \le 28,\cr
0, \mathrm{if}\;k > 28.\cr}
$$
In this case the inverse cascade of wave action was completely suppressed. We observed formation of the direct cascade,
reasonably described by the standard KZ-spectrum $|a_k|^2\simeq\beta_1 k^{-4}$, with $\beta_1 \simeq 1.2\times 10^{-4}$.
This spectrum was observed in the range of scales $32 < k < 150$. The spectra in the linear and logarithmic scales are presented
in Figures~\ref{figure1},\,\ref{figure2}.
The steepness grows in this range approximately as shown in Figure~\ref{figure3}.
At $k_{max}\simeq 200$ steepness reaches its saturation level $\mu_{max}\simeq 0.104$. The shapes of spectral lines at $k$
equal to $50$, $100$, $150$, $200$, and $250$ are represented in Figures~\ref{figure4}-\ref{figure8}.
One can see that the
spectral lines are narrow. Only at area of large $k\simeq 250$ we observe bounded (slave) harmonics which do not obey
linear dispersion relation.
\clearpage

\subsection{Without condensate, inverse cascade is present}
\begin{figure}[tb]
%\vspace*{2mm}
\begin{center}
\includegraphics[width=8.3cm]{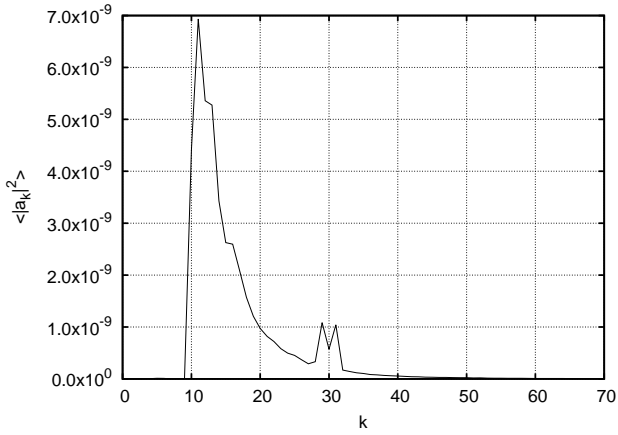}
\end{center}
\caption{\label{figure9}Spectrum with condensate suppressed. Linear scale.}
\end{figure}

\begin{figure}[tb]
%\vspace*{2mm}
\begin{center}
\includegraphics[width=8.3cm]{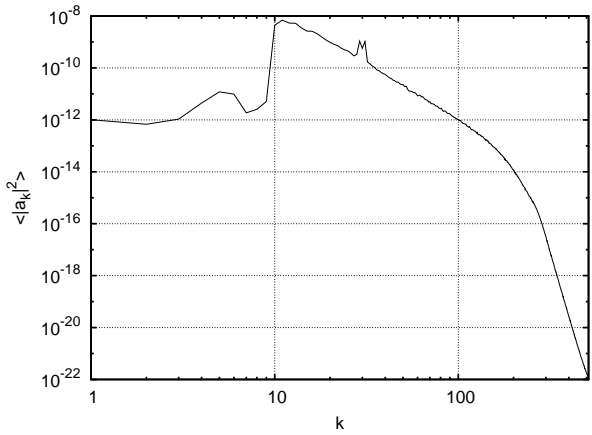}
\end{center}
\caption{\label{figure10}Spectrum with condensate suppressed. Logarithmic scale.}
\end{figure}

\begin{figure}[tb]
%\vspace*{2mm}
\begin{center}
\includegraphics[width=8.3cm]{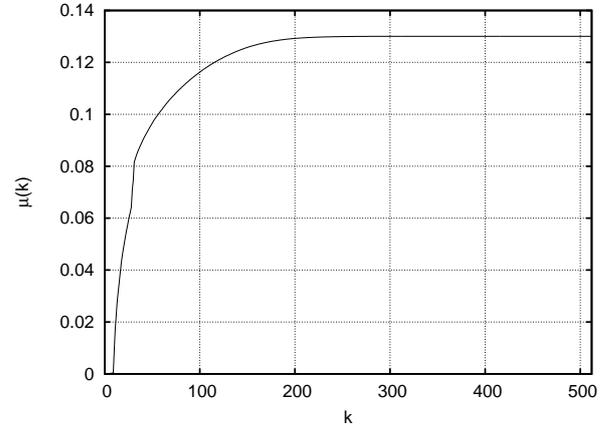}
\end{center}
\caption{\label{figure11}Steepness as a function of absolute vale of wave-number. Only condensate was suppressed.}
\end{figure}

\begin{figure}[tb]
%\vspace*{2mm}
\begin{center}
\includegraphics[width=8.3cm]{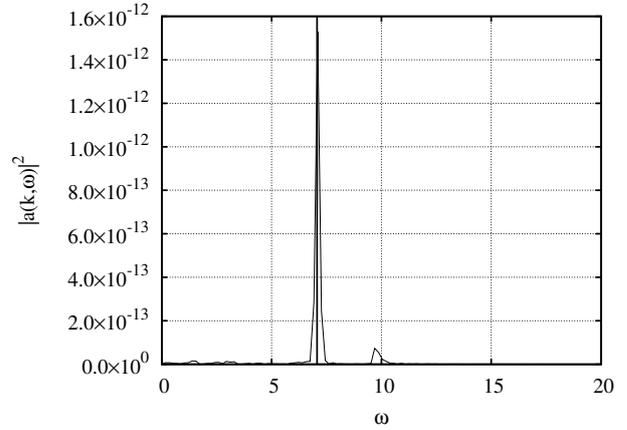}
\end{center}
\caption{\label{figure12}Spectral line of the harmonic $\vec k = (0, 50)^{T}$. Only condensate was suppressed.}
\end{figure}

\begin{figure}[tb]
%\vspace*{2mm}
\begin{center}
\includegraphics[width=8.3cm]{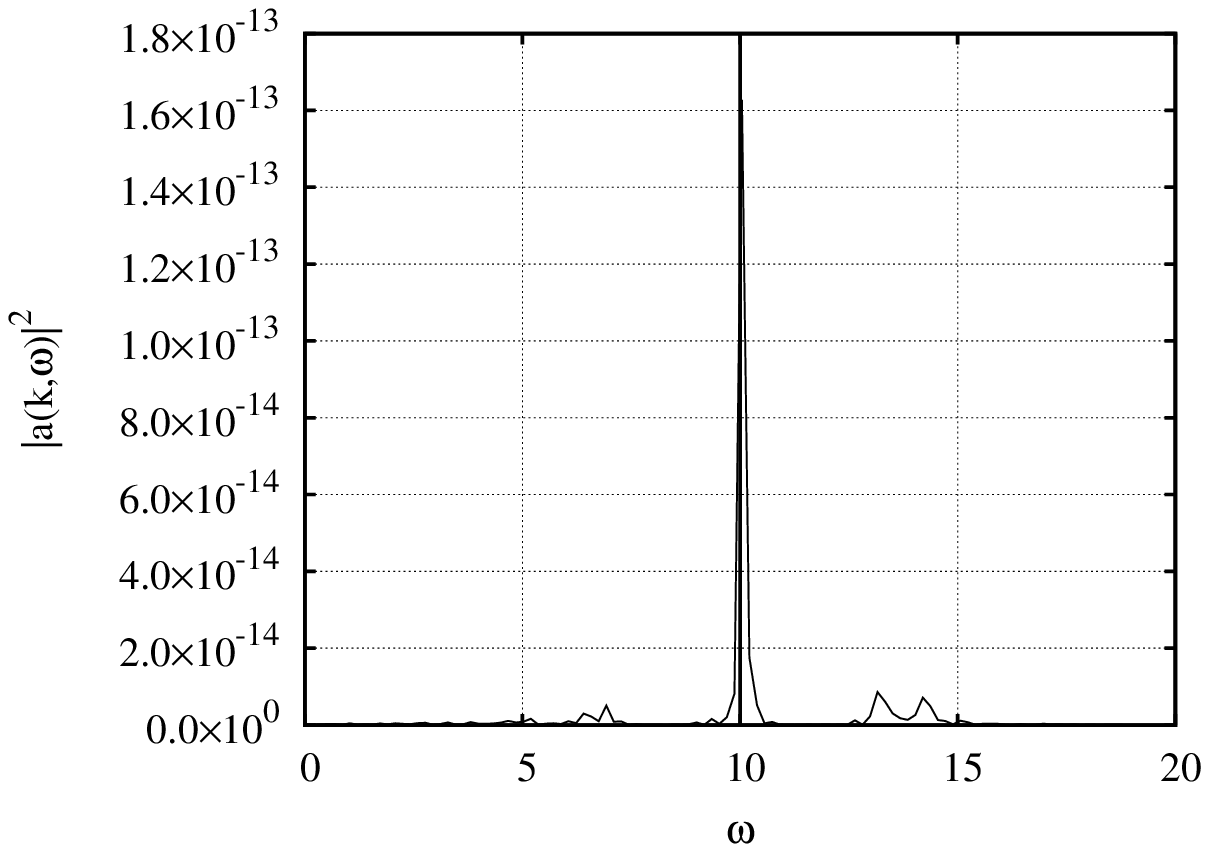}
\end{center}
\caption{\label{figure13}Spectral line of the harmonic $\vec k = (0, 100)^{T}$. Only condensate was suppressed.}
\end{figure}

\begin{figure}[tb]
%\vspace*{2mm}
\begin{center}
\includegraphics[width=8.3cm]{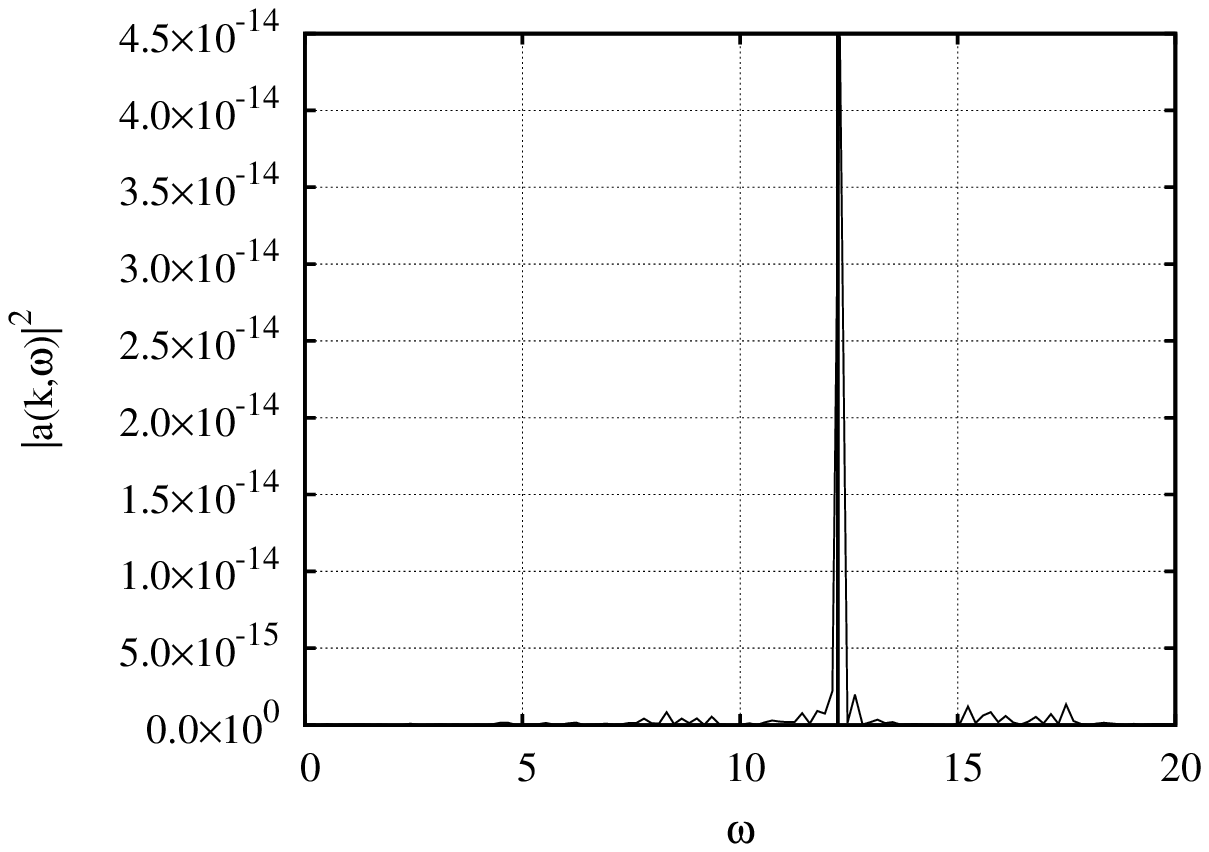}
\end{center}
\caption{\label{figure14}Spectral line of the harmonic $\vec k = (0, 150)^{T}$. Only condensate was suppressed.}
\end{figure}

\begin{figure}[tb]
%\vspace*{2mm}
\begin{center}
\includegraphics[width=8.3cm]{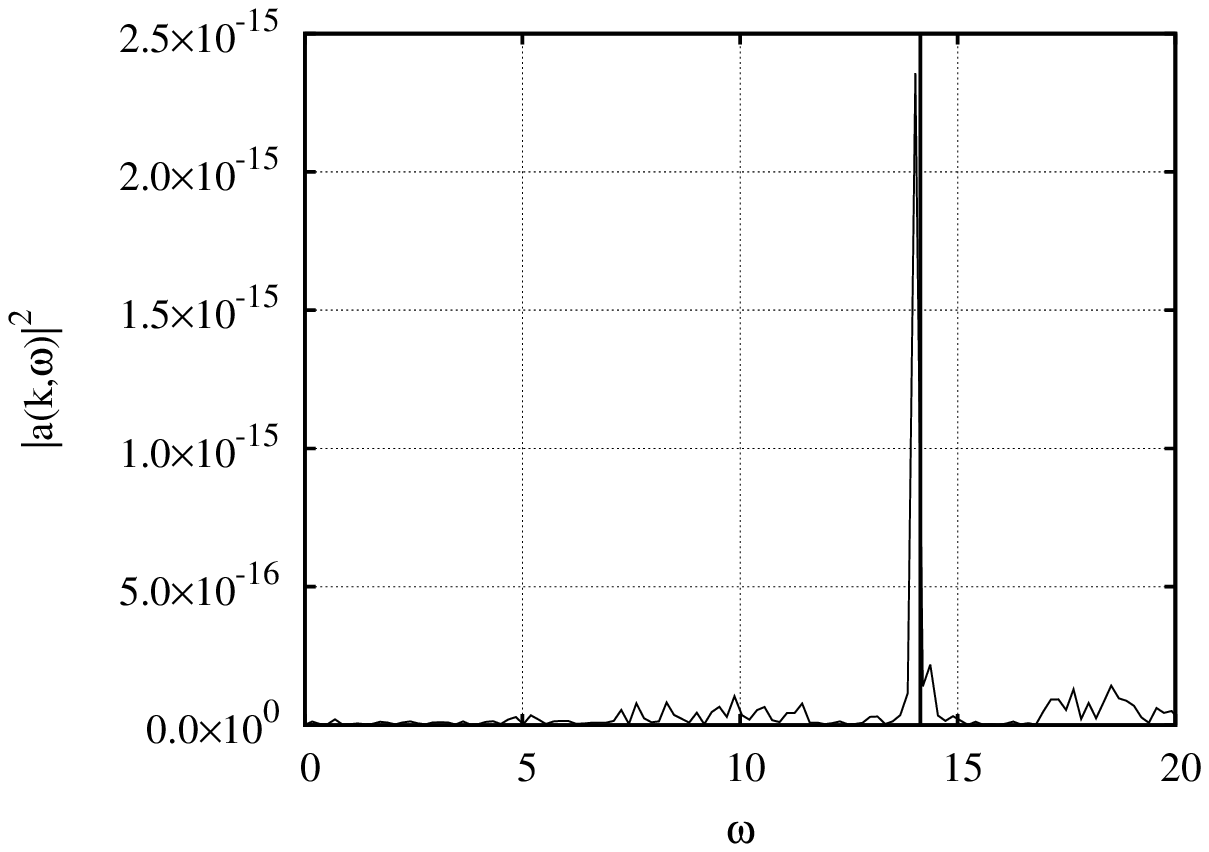}
\end{center}
\caption{\label{figure15}Spectral line of the harmonic $\vec k = (0, 200)^{T}$. Only condensate was suppressed.}
\end{figure}

\begin{figure}[tb]
%\vspace*{2mm}
\begin{center}
\includegraphics[width=8.3cm]{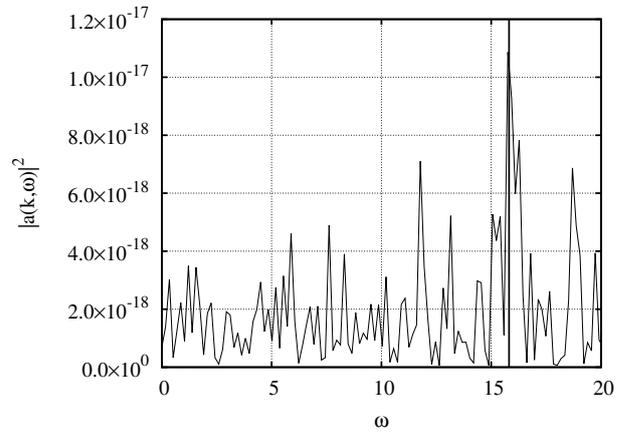}
\end{center}
\caption{\label{figure16}Spectral line of the harmonic $\vec k = (0, 250)^{T}$. Only condensate was suppressed.}
\end{figure}

In the second series of experiments the low frequency damping was presented only in small wave numbers $k<10$, where
the dissipation rate was constant $\gamma_k^{(2)} = 0.05$. In this experiment we observed formation of both direct
and inverse cascades. The dynamic range for inverse cascade was too short for accurate evaluation of the power of the spectrum,
the only statement which can be made is that it is relatively close to the predicted $|a_{\vec k}|^2\sim k^{-23/6}$
spectrum, with power exponent $\alpha$ in $k^{-\alpha}$ ranging from $3.1$ to $4.0$. In the area of direct cascade we
observed slightly steeper spectrum than KZ-spectrum slope. In the area $32<k<150$ the spectrum can be approximated
by $|a_{\vec k}|^2 \sim k^{-4.2}$. This deviation from pure KZ-slope $k^{-4}$ can be explained by the influence of
wave breaking and white capping effects~\citep{Korotkevich2012MCS}.
The spectra are presented in Figures~\ref{figure9},\,\ref{figure10}.
The steepness
in this case reaches the level $\mu\simeq 0.130$ (see Figure~\ref{figure11}).
The spectral lines in this series of
experiments were still narrow for $k=50, 100, 150, 200$. In the area of significant damping $k=250$ we observed
intensive formation of slave (bond) harmonics (see Figures~\ref{figure12}-\ref{figure16}).
\clearpage

\subsection{With both condensate and inverse cascade}
\begin{figure}[tb]
%\vspace*{2mm}
\begin{center}
\includegraphics[width=8.3cm]{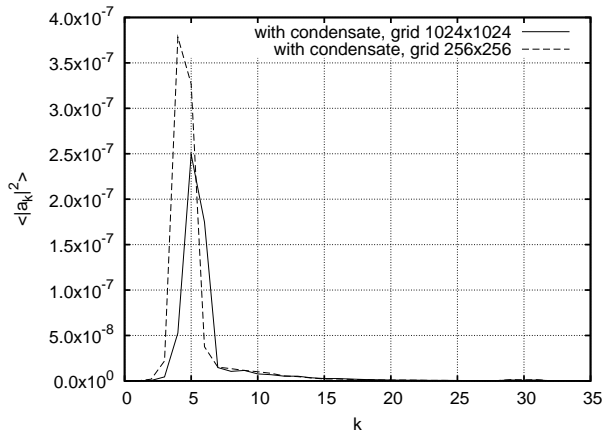}
\end{center}
\caption{\label{figure17}Spectrum with condensate. Two grid sizes. Linear scale.}
\end{figure}

\begin{figure}[tb]
%\vspace*{2mm}
\begin{center}
\includegraphics[width=8.3cm]{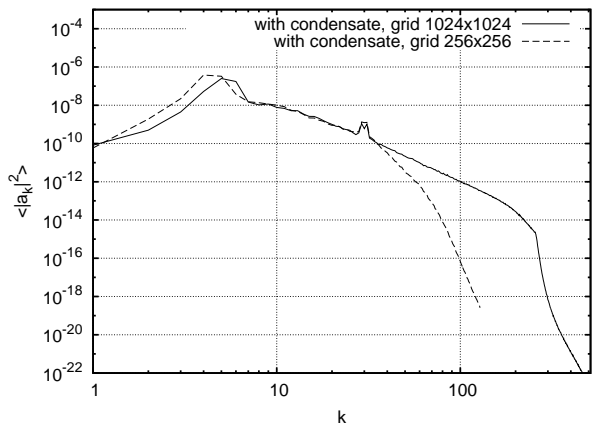}
\end{center}
\caption{\label{figure18}Spectrum with condensate. Two grid sizes. Logarithmic scale.}
\end{figure}

\begin{figure}[tb]
%\vspace*{2mm}
\begin{center}
\includegraphics[width=8.3cm]{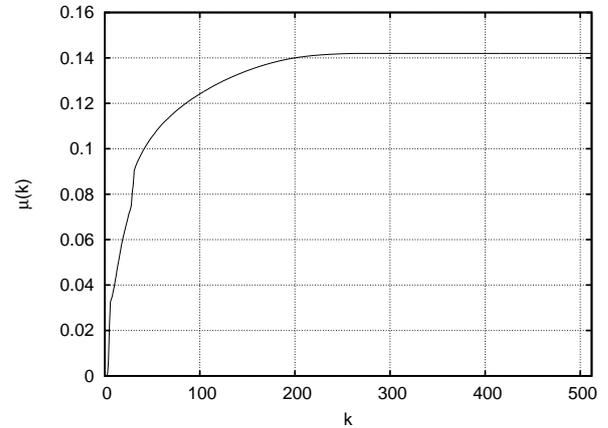}
\end{center}
\caption{\label{figure19}Steepness as a function of absolute vale of wave-number. Condensate was present.}
\end{figure}

\begin{figure}[tb]
%\vspace*{2mm}
\begin{center}
\includegraphics[width=8.3cm]{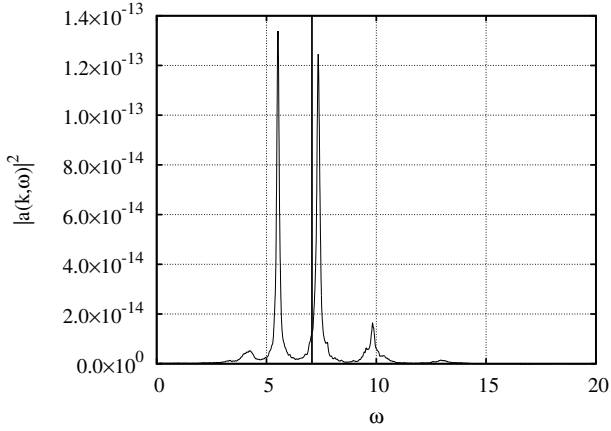}
\end{center}
\caption{\label{figure20}Spectral line of the harmonic $\vec k = (0, 50)^{T}$. Condensate was present.}
\end{figure}

\begin{figure}[tb]
%\vspace*{2mm}
\begin{center}
\includegraphics[width=8.3cm]{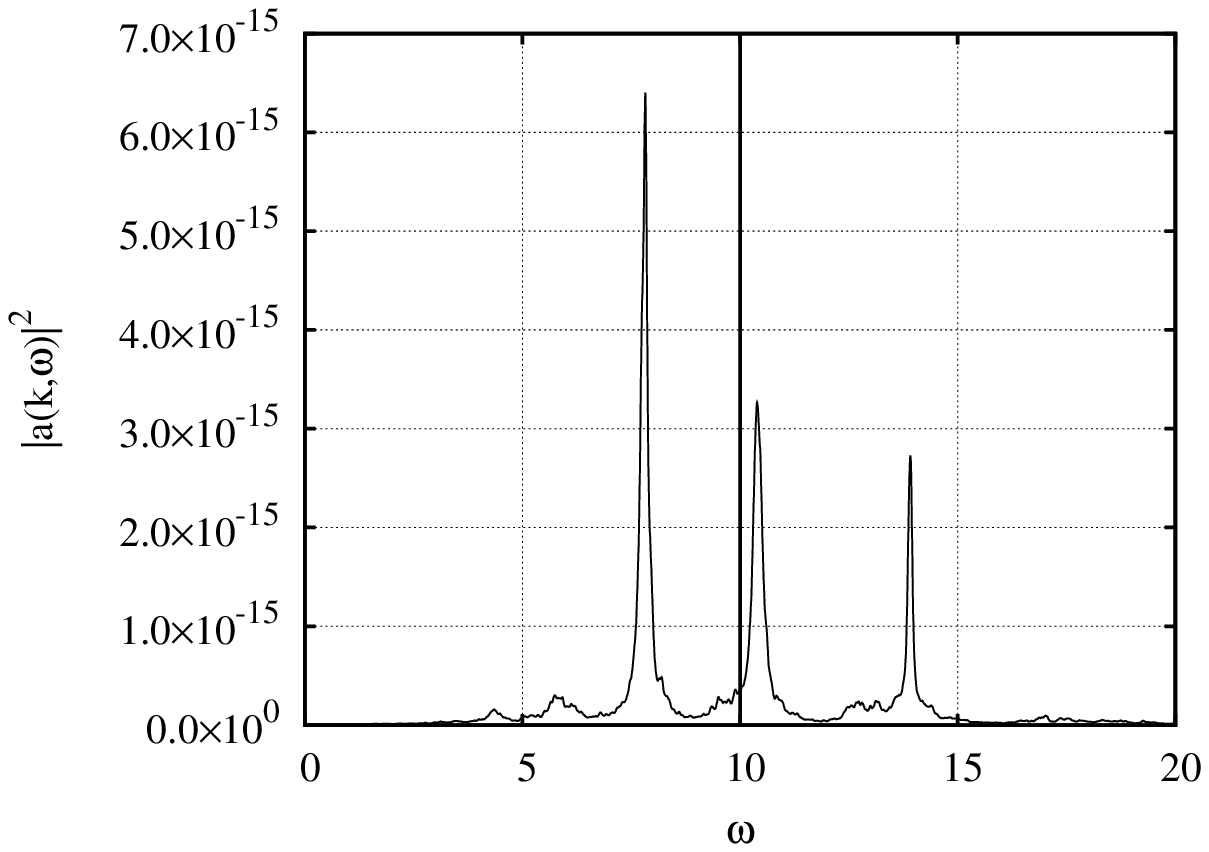}
\end{center}
\caption{\label{figure21}Spectral line of the harmonic $\vec k = (0, 100)^{T}$. Condensate was present.}
\end{figure}

\begin{figure}[tb]
%\vspace*{2mm}
\begin{center}
\includegraphics[width=8.3cm]{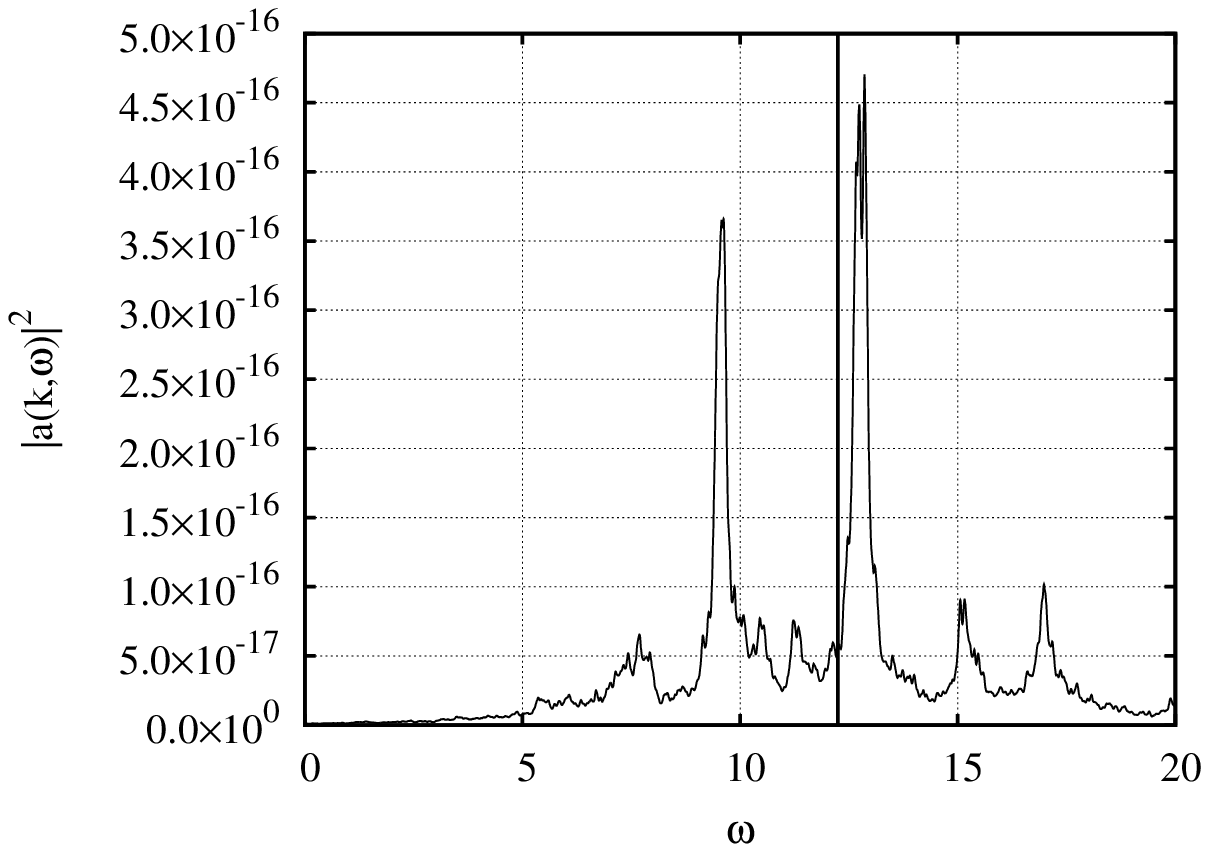}
\end{center}
\caption{\label{figure22}Spectral line of the harmonic $\vec k = (0, 150)^{T}$. Condensate was present.}
\end{figure}

\begin{figure}[tb]
%\vspace*{2mm}
\begin{center}
\includegraphics[width=8.3cm]{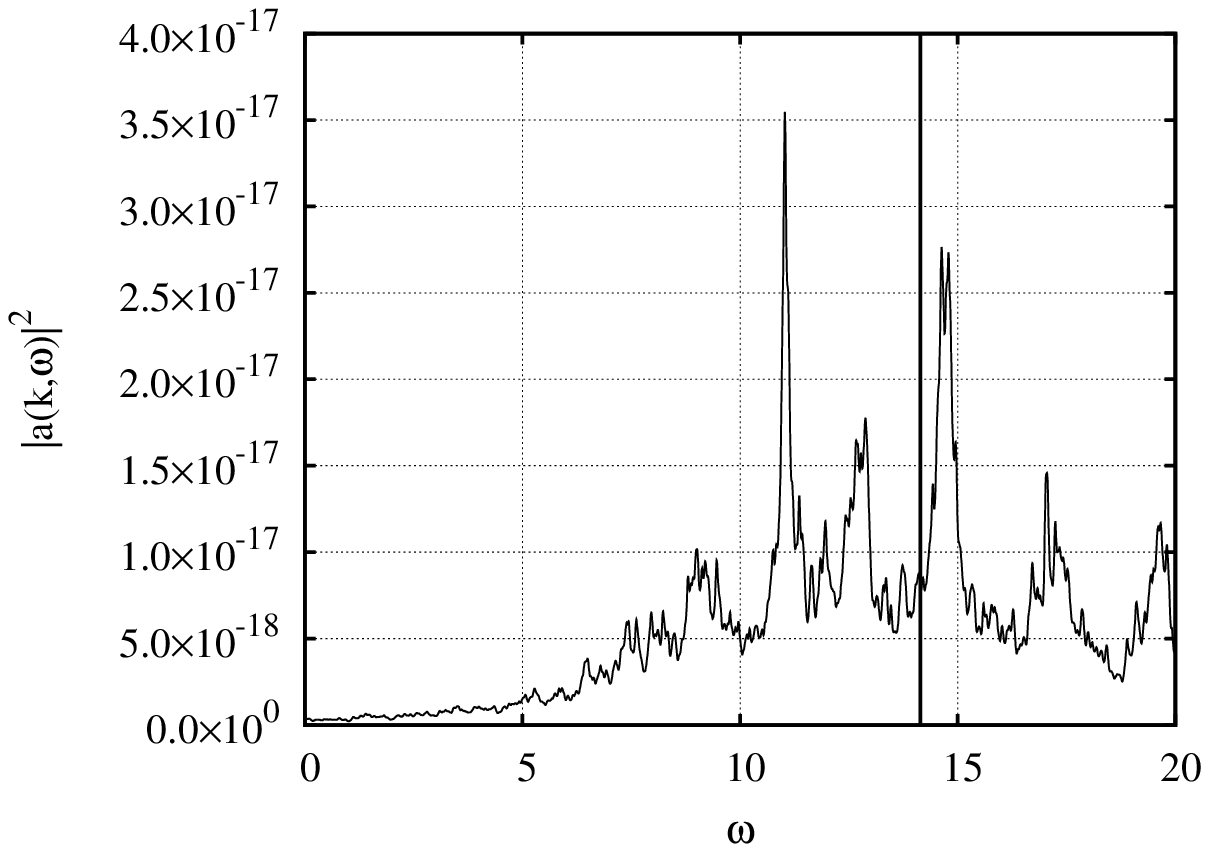}
\end{center}
\caption{\label{figure23}Spectral line of the harmonic $\vec k = (0, 200)^{T}$. Condensate was present.}
\end{figure}

\begin{figure}[tb]
%\vspace*{2mm}
\begin{center}
\includegraphics[width=8.3cm]{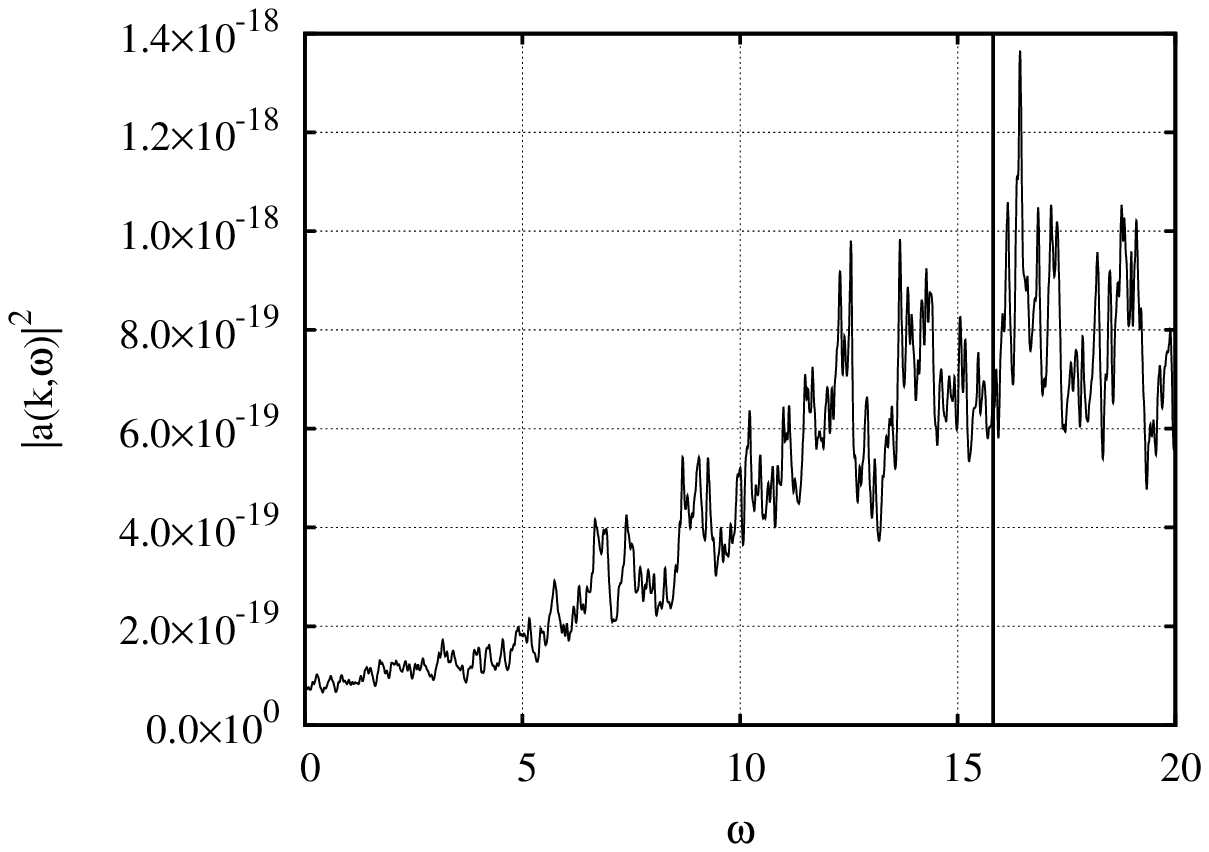}
\end{center}
\caption{\label{figure24}Spectral line of the harmonic $\vec k = (0, 250)^{T}$. Condensate was present.}
\end{figure}

In the last series of experiments we completely eliminated low-frequency dissipation $\gamma_k^{(2)}=0$.
In this case we observed formation of intensive inverse cascade leading to creation of the condensate at $k\simeq 5-4$.
The spectrum in the area of inverse cascade was the same as in the previous series of experiments, while in the area
of direct cascade we observed the Phillips spectrum $|a_{\vec k}|^2\sim k^{-9/2}$ instead of a weak-turbulent
KZ-spectrum~\citep{Korotkevich2008PRL}.
The spectra are presented in Figures~\ref{figure17},\,\ref{figure18}.
In order to check that spectrum slope is not changing any more, long time calculation was performed on a smaller grid,
which showed motion of the condensate position by one wave-number grid step without any significant influence on the
inverse cascade region.
The average steepness was essentially higher than in
the previous experiments reaching its ``limiting value'' $\mu\simeq 0.142$ (see Figure~\ref{figure19}).
We would like to stress that the difference with the previous case, which looks quite small in absolute value
($0.130$ and $0.142$), is really quite strong, because the dissipation rate depends strongly on the average steepness
in this range of values~\citep{ZKP2009} as well as probability of white capping grows pretty fast~\citep{BBY2000}.
It is important to notice that the steepness of the condensate was quite moderate $\mu\simeq 0.06$ (it is worth to note,
that average steepness getting significant contribution from small scales, quite far from the spectral peak, which
means that often used definition of the average steepness through the product of mean amplitude and wave-number of spectral peak
can deviate significantly from the geometrical definition through the average slope of the surface). However modulation
of the short waves by long waves caused intensive micro-breaking of waves which forms the Phillips spectrum
(the process described qualitatively in~\citet{Korotkevich2008PRL,Korotkevich2012MCS}). The most interesting and important
question is about the shapes of spectral lines in the area of Phillips spectrum. They are presented in
Figures~\ref{figure20}-\ref{figure24}. For this experiment we analysed the longest time series which resulted in higher
frequency resolution. After obtaining the frequency spectrum we used moving averaging in order to get rid of noise.
One can see that in the most interesting area $32<k<150$ the spectral lines are
still narrow while essentially ``contaminated'' by the slave harmonics. The maxima of the spectral peaks are shifted
to the high-frequency area according to the theoretically predicted nonlinear frequency shift.
In the area of shorter waves $k>200$ the spectrum is a chaotic mixture of leading and slave harmonics.

\conclusions  %% \conclusions[modified heading if necessary]
We believe that our experiments make possible a ``marriage of Heaven and Hell'' in spirit of William Blake.
Both outstanding scientists -- O.\,Phillips and K.\,Hasselmann are right. If the local steepness $\mu$ is small
($\mu\le 0.1$) the Hasselmann kinetic equation is valid without any augmentation by any additional dissipation terms.
For $\mu$ significantly higher than $0.1$,
in the area of very short waves the kinetic equation is not applicable in its ``pure'' form. It is impossible
to separate ``leading'' and ``slave'' harmonics in this area. This part of the ocean spectrum cannot be described
analytically by the use of perturbative methods. Nevertheless, the Phillips spectrum in this area is still applicable.
The only theoretical reason for this statement is dimensional considerations supported by experimental data.
At the same time we claim that there is a ``grey area'', which is especially interesting, because it is containing most
of energy and generate most of steepness, where micro-breaking have equal foot with weakly nonlinear
resonant waves interaction. Spectra in this area could be described by an ``augmented'' Hasselmann equation, including
an additional term describing dissipation energy due to wave-breaking. Similar additional ``dissipation terms'' $S_{diss}$
are widely used in well developed operational wave forecasting models. But they are introduced ``out of the blue''
and not supported neither by theoretical consideration nor by experimental observations. Moreover, it is not clear
{\it a priori} that one can use the Hasselmann kinetic equation in a situation where wave-breaking events are frequent
enough. We hope that our experiments showed that this is possible. A necessary and sufficient condition of applicability
of the Hasselmann equation is narrowness of the spectral line. In the present article we assert that in the ``grey area''
with a frequency range in one half of the decade the frequency spectra of harmonics are still narrow lines. It means
that the Hasselmann equation is applicable there. Of course it must be augmented by a proper dissipative term. The existing
and widely used dissipative terms hardly can be correct. They do not satisfy to the minimal requirement -- the Hasselmann
equation in the presence of this term has to have the Phillips spectrum as an exact solution. Hence, the urgent problem
is a construction of a ``really good'' dissipative term. One possible and hopefully plausible variant was recently offered
by~\citet{ZPR2013}. The dependence of the dissipation term on the average steepness was recently measured directly from the
numerical experiment, the preliminary results can be found in~\citet{ZKP2009}. In our next paper we shall analyse
what information about possible shape of the dissipative term can be extracted from massive numerical experiments.

\begin{acknowledgements}
The authors gratefully wish to acknowledge the following contributions:
KAO was supported by the NSF grant 1131791, and during the summer visit by the grant NSh-6885.2010.2.

ZVE was partially supported by the NSF grant 1130450 and by the Grant No. 11.G34.31.0035 of the Government of Russian Federation.

Also authors would like to thank developers of FFTW~\citep{FFTW} and the whole GNU project~\citep{GNU} for developing, and supporting this useful and free software.
\end{acknowledgements}

%\bibliographystyle{copernicus}
%\bibliography{surfacewaves}

\begin{thebibliography}{22}
\providecommand{\natexlab}[1]{#1}
\providecommand{\url}[1]{{\tt #1}}
\providecommand{\urlprefix}{URL }
\expandafter\ifx\csname urlstyle\endcsname\relax
  \providecommand{\doi}[1]{doi:\discretionary{}{}{}#1}\else
  \providecommand{\doi}{doi:\discretionary{}{}{}\begingroup
  \urlstyle{rm}\Url}\fi

\bibitem[{Banner et~al.(2000)Banner, Babanin, and Young}]{BBY2000}
Banner, M.~L., Babanin, A.~V., and Young, I.~R.: Breaking probability for
  dominant waves on the sea surface, J.\,Phys.\,Oceanogr., 30, 3145--3160,
  2000.

\bibitem[{Dias et~al.(2008)Dias, I.Dyachenko, and Zakharov}]{DDZ2008}
Dias, F., I.Dyachenko, A., and Zakharov, V.~E.: Theory of weakly damped
  free-surface flows: A new formulation based on potential flow solutions,
  Phys.\,Lett.\,A, 372, 1297--1302, 2008.

\bibitem[{Dyachenko et~al.(1992)Dyachenko, Newell, Pushkarev, and
  Zakharov}]{DNPZ1992}
Dyachenko, A.~I., Newell, A.~C., Pushkarev, A., and Zakharov, V.~E.: Optical
  turbulence: weak turbulence, condensates and collapsing fragments in the
  nonlinear Schroedinger equation, Physica\,D, 57, 96--160, 1992.

\bibitem[{Dyachenko et~al.(2003{\natexlab{a}})Dyachenko, Korotkevich, and
  Zakharov}]{DKZ2003cap}
Dyachenko, A.~I., Korotkevich, A.~O., and Zakharov, V.~E.: Decay of the
  monochromatic capillary wave, JETP~Lett., 77, 477--481, 2003{\natexlab{a}}.

\bibitem[{Dyachenko et~al.(2003{\natexlab{b}})Dyachenko, Korotkevich, and
  Zakharov}]{DKZ2003grav}
Dyachenko, A.~I., Korotkevich, A.~O., and Zakharov, V.~E.: Weak turbulence of
  gravity waves, JETP~Lett., 77, 546--550, 2003{\natexlab{b}}.

\bibitem[{Dyachenko et~al.(2004)Dyachenko, Korotkevich, and Zakharov}]{DKZ2004}
Dyachenko, A.~I., Korotkevich, A.~O., and Zakharov, V.~E.: Weak turbulent
  Kolmogorov spectrum for surface gravity waves, Phys.\,Rev.\,Lett., 92,
  134\,501, 2004.

\bibitem[{Frigo and Johnson(2005)}]{FFTW}
Frigo, M. and Johnson, S.~G.: The design and implementation of FFTW 3,
  Proc.\,IEEE, 93, 216--231, \urlprefix\url{http://fftw.org}, 2005.

\bibitem[{GNU(1984-2012)}]{GNU}
GNU: \urlprefix\url{http://gnu.org}, 1984-2012.

\bibitem[{Hasselmann(1962)}]{Hasselmann1962}
Hasselmann, K.: On the non-linear energy transfer in a gravity-wave spectrum
  Part 1. General theory, J.\,Fluid\,Mech., 12, 481--500, 1962.

\bibitem[{Korotkevich(2008)}]{Korotkevich2008PRL}
Korotkevich, A.~O.: Simultaneous numerical simulation of direct and inverse
  cascades in wave turbulence, Phys.\,Rev.\,Lett., 101, 074\,504, 2008.

\bibitem[{Korotkevich(2012)}]{Korotkevich2012MCS}
Korotkevich, A.~O.: Influence of the condensate and inverse cascade on the
  direct cascade in wave turbulence, Math.\,Comput.\,Simul., 82, 1228--1238,
  2012.

\bibitem[{Korotkevich et~al.(2008)Korotkevich, Pushkarev, Resio, and
  Zakharov}]{KPRZ2008}
Korotkevich, A.~O., Pushkarev, A., Resio, D., and Zakharov, V.~E.: Numerical
  Verification of the Weak Turbulent Model for Swell Evolution,
  Eur.\,J.\,Mech.\,B/Fluids, 27, 361--387, 2008.

\bibitem[{Korotkevich et~al.(2012)Korotkevich, Dyachenko, and
  Zakharov}]{KDZ2012}
Korotkevich, A.~O., Dyachenko, A.~I., and Zakharov, V.~E.: Numerical simulation
  of surface waves instability on a discrete grid, submitted to
  J.\,Comput.\,Phys., 2012.

\bibitem[{Phillips(1958)}]{Phillips1958}
Phillips, O.~M.: The equilibrium range in the spectrum of wind-generated ocean
  waves, J.\,Fluid\,Mech., 4, 426--434, 1958.

\bibitem[{Zakharov and Filonenko(1967{\natexlab{a}})}]{ZF1967}
Zakharov, V.~E. and Filonenko, N.~N.: Energy Spectrum for Stochastic
  Oscillations of the Surface of a Liquid, Sov.\,Phys.\,Dokl., 11, 881--884,
  1967{\natexlab{a}}.

\bibitem[{Zakharov and Filonenko(1967{\natexlab{b}})}]{ZF1967JAMTP}
Zakharov, V.~E. and Filonenko, N.~N.: Energy Spectrum for Stochastic
  Oscillations of the Surface of a Liquid, J.\,Appl.\,Mech.\,Tech.\,Phys., 4,
  506, 1967{\natexlab{b}}.

\bibitem[{Zakharov and Zaslavskii(1982)}]{ZZ1982}
Zakharov, V.~E. and Zaslavskii, M.~M.: The kinetic equation and Kolmogorov
  spectra in the weak turbulence theory of wind waves,
  Izv.\,Atm.\,Ocean.\,Phys., 18, 747--753, 1982.

\bibitem[{Zakharov et~al.(1992)Zakharov, Lvov, and Falkovich}]{ZLF1992}
Zakharov, V.~E., Lvov, V.~S., and Falkovich, G.: Kolmogorov Spectra of
  Turbulence I, Springer-Verlag, Berlin, 1992.

\bibitem[{Zakharov et~al.(2005)Zakharov, Korotkevich, Pushkarev, and
  Dyachenko}]{ZKPD2005}
Zakharov, V.~E., Korotkevich, A.~O., Pushkarev, A., and Dyachenko, A.~I.:
  Mesoscopic Wave Turbulence, JETP\,Lett., 82, 487--491, 2005.

\bibitem[{Zakharov et~al.(2007)Zakharov, Korotkevich, Pushkarev, and
  Resio}]{ZKPR2007}
Zakharov, V.~E., Korotkevich, A.~O., Pushkarev, A., and Resio, D.: Coexistence
  of weak and strong wave turbulence in a swell propagation,
  Phys.\,Rev.\,Lett., 99, 164\,501, 2007.

\bibitem[{Zakharov et~al.(2009)Zakharov, Korotkevich, and Prokofiev}]{ZKP2009}
Zakharov, V.~E., Korotkevich, A.~O., and Prokofiev, A.~O.: On Dissipation
  Function of Ocean Waves Due to Whitecapping, AIP Proceedings, CP1168, 2,
  1229--1231, 2009.

\bibitem[{Zakharov et~al.(2012)Zakharov, Pushkarev, and Resio}]{ZPR2013}
Zakharov, V.~E., Pushkarev, A., and Resio, D.: New wind input term through
  experimental, theoretical and numerical consideration, Submitted to
  J.\,Phys.\,Ocean., 2012.

\end{thebibliography}

\end{document}